\documentstyle[12pt]{article}

\def\beqa{\begin{eqnarray}}
\def\eeqa{\end{eqnarray}}
\def\beq{\begin{equation}}
\def\eeq{\end{equation}}

\def\ad{\dot{a}}
\def\add{\ddot{a}}

\def\bib#1{$^{\ref{#1}}$}

\let\LAM=\Lambda
\let\eps=\varepsilon
\let\gam=\gamma
\let\alp=\alpha
\let\sig=\sigma
\let\OME=\Omega

\def\pr{{\it Phys. Rev.}\ }
\def\pl{{\it Phys. Lett.}\ }
\def\nat{{\it Nature}\ }
\def\apj{{\it Ap. J.}\ }
\def\mnras{{\it Mon. Not. R. Ast. Soc.}\ }
\def\araa{{\it Ann. Rev. Astr. Ap.}\ }
\def\rmp{{\it Rev. Mod. Phys.}\ }
\def\arns{{\it Ann. Rev. Nucl. Part. Sci.}\ }

\begin{document}
\def\bib#1{[{\ref{#1}}]}
\begin{titlepage}
\title{A time--dependent ``Cosmological Constant'' Fenomenology}

 \author{{S. Capozziello, R. de Ritis, A. A. Marino}\\ 
 {\em {\small Dipartimento di Scienze Fisiche, Universit\`{a} di Napoli,}}\\ 
 {\em {\small Istituto Nazionale di Fisica Nucleare, Sezione di Napoli,}}\\
 {\em {\small Mostra d'Oltremare pad. 19 I-80125 Napoli, Italy.}}}
	      \date{}
	      \maketitle
	      \begin{abstract}
We construct a cosmological toy model in which a step-function 
``cosmological constant'' is taken into consideration beside ordinary
matter. We assume that $\Lambda$ takes two values depending on the
epoch, and matter goes from a radiation dominated era to a dust dominated
era. The model is exactly solvable and it can be compared with 
recent observations.
	      \end{abstract}

\vspace{20. mm}
e-mail address:\\
 capozziello@axpna1.na.infn.it\\
 deritis@axpna1.na.infn.it\\
 marino@axpna1.na.infn.it
	      \vfill
	      \end{titlepage}

\section{\normalsize \bf Introduction}
A cosmological constant term in the energy density of the universe turns out
to be taken in a very serious consideration in the today research. Many
experimental data on the structure of the present universe are compatible
with models in which the cosmological constant term contribution in the
density parameter $\Omega$ (the mean mass density relative to the critical
density of the corresponding Einstein-de Sitter model $\rho_c=
{\displaystyle \frac{3 H^2}{8 \pi G}}$) is relevant with respect to the
matter term. This happens in the number count of galaxies as well as in the
spectrum of the Cosmic Microwave Background Radiation (CMBR)
\bib{fuku-taka}, \bib{ostri}. On one side there is the well known problem
concerning the theoretical and experimental limits on the values the
cosmological costant should assume, in the sense that they differ for 120
order of magnitude \bib{carrol}, \bib{wein}. On the other side, despite of
these discrepancies, the presence of a cosmological constant term is
requested to avoid the problems which are found comparing the age of the
universe as coming from estimates on globular clusters with the value 
obtained from a standard model with an unitary density parameter, if one
wants to give to the Hubble parameter $H_o$ a value in agreement with the
most recent results \bib{ostri}, \bib{fuku-hoga}, \bib{freed}. As it is well
known, the value for the age comes out to be too low, while with a
cosmological constant term one is able to find the agreement with the
observations. Finally from the statistic of the phenomenon of the
gravitational lensing it comes out an upper limit on $\LAM$, that is
its contribution to $\OME$ seems to be less than $95 \%$ \bib{fuku-futa}.

Furthermore the models in which a cosmological constant is present give rise
to the inflationary epoch which solves the problems of the cosmological
standard model such as the horizon, flatness, entropy problems \bib{guth},
\bib{linde}. In such a context, the present value of the density parameter
turns out to be very close to unity. If we consider that the observational
estimates of $\Omega$ from the smaller scales of galaxies to scales of order
of $10~ Mpc$ give $0.05 \leq \OME \leq 0.2$ \bib{peebles}, we come out at
once to the very well known problem of $\Omega$.

In this article we analyse some phenomenological aspects relatively to the
presence of a cosmological constant in connection with the problem of
$\Omega$ in the context of an exactly solved model. We consider a
homogeneous and isotropic flat model with an inflationary epoch, a radiation
dominated epoch and a matter dominated epoch, in which there is a residual
cosmological constant. That is we consider the system of equations
\beq 
\label{1}
2 \frac{\add}{a}+ \left( \frac{\ad}{a} \right)^2= \LAM- p_m
\eeq
\beq 
\label{2}
\dot{\rho}_m+ 3 \frac{\ad}{a} (p_m+ \rho_m)= 0
\eeq
\beq 
\label{3}
\left( \frac{\ad}{a} \right)^2= \frac{\LAM}{3}+ \frac{\rho_m}{3}
\eeq
in which $a(t)$ is the expansion parameter; actually $\Lambda$ is considered
function of time, more precisely, as it will be shown, it will be
considered piecewise constant; $\rho_m$, $p_m$ are the energy density and
the pressure relative to the matter. The eqs. (\ref{1}), (\ref{3}) are the
Einstein equations, while (\ref{2}) is the contracted Bianchi identity. We
are using units in which $8\pi G= c= 1$. Finally, considering standard
matter, its state equation is, as usual, $p_m= (\gam- 1)\rho_m$, with $1
\leq \gam \leq 2$.

\section{\normalsize \bf The Model}

We describe  three epochs in which  we assume $\Lambda$ and
$\gamma$ as step-functions, that is 
\beq
\label{5}
\LAM(t)= \left \{ 
\begin{array}{ll}
\mbox{$\LAM_1,~~$} & \mbox{$t_i < t < t_f$} \\
\mbox{$\LAM_2,~~$} & \mbox{$t_f < t$}
\end{array} \right.
\eeq
\beq
\label{6}
\gam(t)= \left \{
\begin{array}{ll}
\mbox{$\gam_1= {\displaystyle\frac{4}{3}},~~$} & \mbox{$t < t_{eq}$} \\ 
[2.mm] \mbox{$\gam_2= 1,~~$} & \mbox{$t_{eq} < t$}
\end{array} \right.
\eeq
in which $\Lambda_1$ is the value of the cosmological constant during the
inflationary epoch, $\Lambda_2$ is the residual value, $t_i,~ t_f$ are the
initial and final times of the inflationary epoch, $t_{eq}$ is the instant
of equivalence between radiation and matter density. We solve the
(\ref{1})-(\ref{3}) for $t_i < t < t_f$, $t_f < t < t_{eq}$, $t_{eq}
< t$ and then impose the continuity of the expansion parameter $a$ and of
the total energy density $\rho_{tot}= \Lambda+ \rho_m$ in $t_f$, $t_{eq}$;
from (\ref{1})-(\ref{3}) these conditions imply that the Hubble parameter
$H= {\displaystyle\frac{\dot{a}}{a}}$ and the total density parameter
$\Omega_{tot}= {\displaystyle\frac{\rho_{tot}}{3~ H^2}}$ are continuos at
any $t$.

To solve (\ref{1})-(\ref{3}), taking into account the state equation,
we follow this way: we multiply eq. (\ref{3}) by a factor
$\beta_i$, then add it to eq. (\ref{1}), obtaining 
\beq 
\label{7}
2 \frac{d}{dt} \left( \frac{\ad}{a} \right)+ (\beta_i+ 3) 
\left( \frac{\ad}{a} \right)^2- \frac{1}{3} (\beta_i+ 3) \LAM_i- 
\frac{1}{3} [\beta_i- 3(\gamma_i- 1)] \rho_m= 0
\eeq
in which the index $i$ takes into account the different values of $\gamma$
and $\Lambda$ in the different epochs, according to (\ref{5}), (\ref{6}).
Taking $\beta_i= 3(\gamma_i- 1)$, we obtain a second order equation for
$a(t)$ in which it does not appear explicitely the term relative to
$\rho_m$. Eq. (\ref{7}) is a Riccati-type equation in ${\displaystyle
\frac{\ad}{a}}$, which is possible to solve and one finds
\beq
\label{10}
a= a_{\alp} \left[ e^{\frac{3 \gam_i}{2} \sqrt{\frac{\LAM_i}{3}}~ t}- 
c_{\alp} e^{-\frac{3 \gam_i}{2} \sqrt{\frac{\LAM_i}{3}}~ t} 
\right]^{2/3 \gam_i}.
\eeq
From the (contracted) Bianchi identity, eq. (\ref{2}), and from the (0,0)
Einstein eq. (\ref{3}) we get $c_{\alp}= {\displaystyle \frac{M_{\alp}}{4
\LAM_i a_{\alp}^{3 \gam_i}}}$, where $M_{\alpha}$ is given by $M_{\alp}=
\rho_m a^{3 \gam_i}$.

It is interesting to note that the expression (\ref{10}) for $a(t)$ presents 
a singularity for any ($\alpha$, $i$) if $c_{\alpha} > 0$. This condition is
verified, being $c_{\alpha}$ connected with the energy density of matter.
Thus there will be a $t_s$ such that $a(t_s)= 0$ and it
seems natural to put the time origin in $t_s$. Redefining $t$ such that
$\left. a \right|_{t=0}= 0$, the solution takes the form 
\beq
\label{13}
a= \left \{
\begin{array}{lll}
\mbox{$a_1 \left[ e^{2 \sqrt{\frac{\LAM_1}{3}}~ t}- 
e^{-2 \sqrt{\frac{\LAM_1}{3}}~ t} \right]^{1/2},~~$} & 
\mbox{$t_i < t < t_f,~~$} & \mbox{$
{\displaystyle\frac{M_1}{4~ \LAM_1~ a_1^4}}= 1;$} \\
\mbox{$a_2 \left[ e^{2 \sqrt{\frac{\LAM_2}{3}}~ t}- 
c_2 e^{-2 \sqrt{\frac{\LAM_2}{3}}~ t} \right]^{1/2},~~$} &
\mbox{$t_f < t < t_{eq},~~$} & \mbox{$c_2= 
{\displaystyle\frac{M_2}{4~ \LAM_2~ a_2^4}};$} \\
\mbox{$a_3 \left[ e^{\frac{3}{2} \sqrt{\frac{\LAM_2}{3}}~ t}- 
c_3 e^{-\frac{3}{2} \sqrt{\frac{\LAM_2}{3}}~ t} \right]^{2/3},~~$} &
\mbox{$t_{eq} < t,~~$} & \mbox{$c_3= 
{\displaystyle\frac{M_3}{4~ \LAM_2~ a_3^3}};$}
\end{array} \right.
\eeq
in which $a_{\alpha}$, $c_{\alpha}$ are not the same as given in (\ref{10})
but they have been opportunely redefined, because of the fixing of 
the time beginning. Of course, the meaning of the
times $t_i$, $t_f$, $t_{eq}$ of the (\ref{5}), (\ref{6}) coincide with
those of (\ref{13}). For the total energy density one has the expression for
the different epochs: 
\beq
\label{14}
\rho_{tot}= \left \{
\begin{array}{ll}
\mbox{$\LAM_1+ {\displaystyle\frac{M_1}{a^4}},~~$} & 
\mbox{$t_i < t < t_f$} \\ [3.mm]
\mbox{$\LAM_2+ {\displaystyle\frac{M_2}{a^4}},~~$} & 
\mbox{$t_f < t < t_{eq}$} \\ [3.mm]
\mbox{$\LAM_2+ {\displaystyle\frac{M_3}{a^3}},~~$} &  
\mbox{$t_{eq} < t$}
\end{array} \right.
\eeq
As we have said above, we impose the continuity conditions
\beq
\label{15}
\left. a\right|_{t_f^-}= \left. a\right|_{t_f^+},~~ 
\left. \rho_{tot} \right|_{t_f^-}= \left. \rho_{tot} \right|_{t_f^+}
\eeq
\beq
\label{16}
\left. a\right|_{t_{eq}^-}= \left. a\right|_{t_{eq}^+},~~ 
\left. \rho_{tot} \right|_{t_{eq}^-}= \left. \rho_{tot} \right|_{t_{eq}^+}.
\eeq
In this way, from (\ref{3}) and from the definition of $\Omega$, we have 
that $H(t)$ and $\Omega(t)$ are continuos at any $t$. 

It is noteworthy that, assuming (\ref{5}), (\ref{6}) and taking into account
(\ref{14}), we introduce a discontinuity, at the instants $t_f$, $t_{eq}$, in
the equation of state \bib{ellis} for the total energy density $p_{tot}=
(\gam_{tot}-1)~ \rho_{tot}$, being $\gam_{tot}= {\displaystyle
\frac{\gam(t)~ \rho_m}{\LAM(t)+ \rho_m}}$; then the entropy $S$ and the
scalar curvature $R$ are discontinuous too in the same instants, being
respectively 
\beq
\label{19}
S= \frac{a^3~ \gam_{tot}}{T};~~ 
R= -6 \left( \frac{\add}{a}+ H^2 \right), 
\eeq
and considering that (\ref{1}) can be written as 
\beq
\label{21}
2 \left(\frac{\add}{a}\right)+ 
\left( \frac{\ad}{a}\right)^2= -(\gam_{tot}- 1) \rho_{tot}.
\eeq
Thus the entropy production in $t_f$ takes place through a phase transition
and takes into account the production of matter at the end of the
inflationary epoch through the condition (\ref{15}) on $\rho_{tot}$. 

Referring to the first of (\ref{14}), we can consider the beginning
of  inflationary epoch $t_i$ as the instant in which there is equivalence
between the energy density relative to the cosmological constant and the one 
relative to the matter, that is 
\beq
\label{22}
\LAM_1= \left. \frac{M_1}{a^4} \right|_{t_i}
\eeq
so that, for $t \ll t_i$, one would have a radiation dominated pre-inflation epoch
and for $t \gg t_i$ one has the inflationary epoch dominated by the
cosmological constant. Thus 
\beq
\label{23}
t_i= \frac{1}{2} \sqrt{\frac{3}{\LAM_1}}~ ln(1+ \sqrt{2}).
\eeq
To solve the problems of the standard model (horizon, flatness, entropy),
the number of e-folding during inflation has to be $N_{e-folding} \geq 67$
\bib{guth}, \bib{linde}. From this condition assuming that
$\sqrt{\Lambda_1/3}~ t \gg 1$ during inflation, which
seems to be very reasonably, we get a condition on $\Lambda_1$ and $t_f$,
given by $\LAM_1~ t_f^2 \geq 1.4 \cdot 10^4$. The validity of the model is
assumed, of course, for $t \geq t_P$, with $t_P \simeq 5.4 \cdot 10^{-44}~
s$ the Planckian time. Imposing that at $t_P$ the total energy density is
just the Planck density, we obtain an estimate for $\Lambda_1$. Thus we write
\beq
\label{25}
\left. \frac{M_1}{a^4} \right|_{t_P}= \rho_P
\eeq
in which $\rho_P$ is the Planckian density. If we assume that at $t_P$ the
radiation dominates, that is $t_P \ll t_i$, and we can develop the first of
(\ref{10}) in $\sqrt{\LAM_1/3}~ t$ to the first
order (radiation behavior for $a(t)$), obtaining at once an estimates for
$\LAM_1$, which comes out to be negative. This means that we have to take
into account also the cosmological constant term. Assuming in the first
approximation that it dominates, we get a lower bound for $\Lambda_1$, that
is $\LAM_1 \geq 8.4 \cdot 10^{87}~ s^{-2}$. From (\ref{23}) one gets $t_i
\leq 8.3 \cdot 10^{-45}~ s$, which is less than the Planckian time. This
only means that our model can be considered reasonable starting with a
cosmological constant dominated era.

The continuity conditions (\ref{15}), (\ref{16}), give  some relations
between the constant $a_{\alpha}$, $c_{\alpha}$ present in eq. (\ref{13});
taking into account the relation between $\LAM_1,~ t_f$ we get 
\beq
\label{28}
a_1= \left[ \frac{2}{1+ \sqrt{\LAM_{12}}}\right]^{1/2}
\frac{ a_o~ e^{\sqrt{\LAM_2/3}~ (t_f- t_o)} 
[ 1- r_{12}~ e^{ 4 \sqrt{\LAM_2/3}~ (t_f- t_{eq})}]^{1/6}}
{ e^{\sqrt{\LAM_1/3}~ t_f} [ 1- r_{12}~ e^{ \sqrt{\LAM_2/3}~ 
(4t_f- t_{eq}- 3t_o)}]^{2/3}} \\
\eeq
\beq
\label{29}
a_2= \frac{ a_o~ 
[ 1- r_{12}~ e^{ 4 \sqrt{\LAM_2/3}~ (t_f- t_{eq})}]^{1/6}}
{ e^{\sqrt{\LAM_2/3}~ t_o} [ 1- r_{12}~ e^{ \sqrt{\LAM_2/3}~ 
(4t_f- t_{eq}- 3t_o)}]^{2/3}} \\
\eeq
\beq
\label{30}
a_3= \frac{ a_o}
{ e^{\sqrt{\LAM_2/3}~ t_o} [ 1- r_{12}~ e^{ \sqrt{\LAM_2/3}~ 
(4t_f- t_{eq}- 3t_o)}]^{2/3}} \\
\eeq
\beq
\label{32}
c_2= r_{12}~ e^{4 \sqrt{\LAM_2/3}~ t_f},~~ 
c_3= r_{12}~ e^{\sqrt{\LAM_2/3}~ (4t_f- t_{eq})} 
\eeq
in which $t_o$ is the present age, $a_o= \left. a\right|_{t_o}$,
${\displaystyle \LAM_{12}= \sqrt{\frac{ \LAM_1}{ \LAM_2}}}$ and $r_{12}
\simeq {\displaystyle \frac{-1+ \sqrt{\frac{ \LAM_1}{ \LAM_2}}} {1+
\sqrt{\frac{ \LAM_1}{ \LAM_2}}}}$.

A very reasonable assumption is $\Lambda_1 > \Lambda_2$; it can be seen
that this is equivalent to have $a_1$, $a_2$, $a_3$ real and $c_2$, $c_3$
positive.

\section{\normalsize \bf Compatibility with observations}

The comparison with the experimental data can be done through the constants
$M_2$, $M_3$, which can be connected with the energy density of the matter
and of the radiation at the present age, which we call respectively $\mu_o$,
$\varepsilon_o$.

Assuming $\varepsilon_o \ll \mu_o$, one can write
\beq
\label{35}
M_2= \eps_o~ a_o^4,~~ M_3= \mu_o~ a_o^3.
\eeq
Using the relations existing between $c_2$, $M_2$ and $c_3$, $M_3$ as given by
(\ref{13}) and taking into account eqs. (\ref{28})-(\ref{32}), we get 
\beq
\label{36}
\eps_o= \frac{4~ \LAM_2~ c_2~ a_2^4}{a_o^4}= 
\frac{ 4~ \LAM_2~  r_{12}~ e^{\sqrt{\LAM_2/3}~ (t_f- t_o)} 
[ 1- r_{12}~ e^{ 4 \sqrt{\LAM_2/3}~ (t_f- t_{eq})}]^{2/3}}
{[ 1- r_{12}~ e^{ \sqrt{\LAM_2/3}~ (4t_f- t_{eq}- 3t_o)}]^{8/3}} 
\eeq
\beq
\label{37}
\mu_o= \frac{4~ \LAM_2~ c_3~ a_3^3}{a_o^3}= 
\frac{ 4~ \LAM_2~  r_{12}~ e^{\sqrt{\LAM_2/3}~ (4t_f- t_{eq}- 3t_o)}}
{[ 1- r_{12}~ e^{ \sqrt{\LAM_2/3}~ (4t_f- t_{eq}- 3t_o)}]^2}.
\eeq
Being $t_f$, $t_{eq} \ll t_o$ we also get an expression for the present 
age
\beq
\label{38}
t_o= \frac{1}{H_o} \int_1^{\infty} \frac{dx}
{x \sqrt{\OME_{\mu_o}~ x^3+ \OME_{\LAM_o}}}
\eeq
in which $H_o$, $\Omega_{\mu_o}$, $\Omega_{\Lambda_o}$ are respectively the
Hubble parameter, the contribution of  matter and  cosmological
constant to $\Omega$ at the present age. The observational value of
$\varepsilon_o$ comes from the black body law 
\beq
\label{39}
\eps_o=  \sig_B~ T_o^4 \simeq 4.649 \cdot 10^{-34}~ g~ cm^{-3}
\eeq
in which $\sigma_B$ is the radiation-density constant and $T_o= 2.726 \pm 
0.010~^oK$ is the CMBR temperature \bib{wright}.

The most recent estimates of  Hubble parameter $H_o$ come from various
distances calibrators on different scales and are (see \bib{ostri}, 
\bib{fuku-hoga}, \bib{freed}, \bib{pier}) 
\beq
\label{40}
H_o= 100~ h~ Km~ s^{-1}~ Mpc^{-1},~~ 0.55 \leq h \leq 0.85.
\eeq
The most recent estimates of $\Omega_{\mu_o}$ come from comparison of CDM
models vs. redshift surveys and from studies on the dynamics of cosmic
flows; they give (see \bib{efsta}, \bib{dekel}, \bib{prim}) $\OME_{\mu_o}
\geq 0.2 \div 0.3$ at $2\sig$ level or more. Thus, from the definition of
$\Omega$, being the model flat, we
get 
\beq
\label{42}
\mu_{o,~ est.}= 3~ H_o^2~ \OME_{\mu_o}= 
1.879 \cdot 10^{-29}~ h^2~ \OME_{\mu_o}~ g~ cm^{-3}
\eeq
\beq
\label{43}
\LAM_{2,~ est.}= 3~ H_o^2~ \OME_{\LAM_o}= 
3.151 \cdot 10^{-35}~ h^2~ \OME_{\LAM_o}~ s^{-2}.
\eeq
Comparing (\ref{42}) with (\ref{39}) we see that the assumption
$\varepsilon_o \ll \mu_o$ is justified. 

We obtain an estimate of $t_f$ considering that, for $t_f < t <
t_{eq}$, the matter is prevalently ultrarelativistic, thus one has the
relation between the energy density $\rho_m$ and the absolute temperature 
$T$ 
\beq
\label{44}
\rho_m= \frac{\sig_B}{2}~ g(T)~ T^4
\eeq
in which $g(T)$ are the effective spin degree of freedom. Considering the
epoch just after $t_f$, the radiation dominates on the cosmological constant
and, being verified by construction the hypothesis of an efficient reheating,
equating the expression of $\rho_m(t_f^+)$ with $\rho_m(T_f^+)$, one gets
$t_f$ in terms of $T_f^+$, that is, the right limit on $T_f$. Such a
quantity  is  constrained by the bariogenesis \bib{kolb}, being $T_f^+ \geq
10^{10}~ GeV$, which implies a constraint on $t_f$, that is $t_f \leq 3.7
\cdot 10^{-29}~ s$ and this is compatible with the relation between $\LAM_1$, 
$t_f$ and with the constraint we found on $\LAM_1$.

Giving to $t_f$, $t_{eq}$, $\OME_{\mu_o}$ respectively the values
$t_f= 10^{-29}$, $t_{eq}= 10^{12}$, $\OME_{\mu_o}= 0.3$,
compatible with all the considerations we have done, we find that the values
of $\mu_o$, $\varepsilon_o$ given from the model are substantially compatible
with those coming from the observations. 

In particularly we find that the values of $\eps_{o,~ mod.}$ obtained from
the model, that is from (\ref{36}), differ from the values $\eps_{o,~ est.}$
given from (\ref{39}) when $h$ varies according with (\ref{40}) as 
\beq
\label{45}
\left. \frac{\eps_{o,~ est.}- \eps_{o,~ mod.}}{\eps_{o,~
est.}} \right|_{h= 0.55}= 0.43,~~ 
\left. \frac{\eps_{o,~ est.}-
\eps_{o,~ mod.}}{\eps_{o,~ est.}} \right|_{h= 0.85}= -0.83,
\vspace{2.mm}
\eeq
essentially unchanged for $T_o$ in the experimental errors and for increasing
$\OME_{\mu_o}$. This means that there is full compatibility; the value of
$h$ which minimizes the square of the relative difference for $\eps_o$ to a
value significantly lower than $10^{-4}$ is given by $h= 0.6777$, which is
compatible with the estimate given by (\ref{40}).

The values of $\mu_{o,~ mod.}$ obtained from the model, that is from
(\ref{37}), differ from the values $\mu_{o,~ est.}$ given from (\ref{42})
when $h$ varies according with (\ref{40}) as 
\beq
\label{46}
\left. \frac{\mu_{o,~ est.}- \mu_{o,~ mod.}}{\mu_{o,~ 
est.}} \right|_{h= 0.55}= 1.8 \cdot 10^{-6},~~ 
\left. \frac{\mu_{o,~ est.}- \mu_{o,~ mod.}}{\mu_{o,~ 
est.}} \right|_{h= 0.85}= 2.2 \cdot 10^{-6}.
\vspace{2.mm}
\eeq
In this case we don't find a full compatibility but the relative difference
is of one part over $10^{6}$. The agreement increases for increasing
$\OME_{\mu_o}$ and decreasing $h$. 

The values of the equivalence temperature, for $h= 0.6777$, are found to be
$\left. T_{eq} \right|_{T_o= 2.716}= 1.303~ eV$, $\left. T_{eq}
\right|_{T_o= 2.736}= 1.313~ eV$, which are just one order of magnitude
less than the decoupling temperature, given by $T_{dec}= 13~ eV$, and for the
temperature immediately after inflation $\left. T_f^+ \right|_{T_o= 2.716}=
5.041 \cdot 10^{18}~ GeV$, $\left. T_f^+ \right|_{T_o= 2.736}= 5.078 \cdot
10^{18}~ GeV$, which are compatible with the constraint imposed by the
bariogenesis.

Moreover we find for the present age the value $t_{o,~ mod.}= 13.9 \cdot
10^9~ y$, for $h= 0.6777$, compatible with the most recent estimates which
give $t_{o,~ est.}= 14 \pm 2 \cdot 10^9~ y$ \bib{freed}. Of course the
agreement decreases with increasing $h$ and increases with decreases
$\OME_{\mu_o}$.

Finally we want to stress that the behavior of $a(t)$ given from (\ref{13}),
seems to be quite different from the standard expansion $a \propto t^{2/3}$
of an universe made of dust, as our present universe appears to be. One gets
the standard expansion from (\ref{13}) if $\sqrt{\LAM_2/3}~ t \ll 1$; this
turns out to be no longer strictly verified at the present age, since we get
$\sqrt{\LAM_2/3}~ t_o \simeq 0.7$. This means that the effects of
cosmological constant term on the expansion are no longer negligible. This
is something which could be taken into consideration in the experimental
measurements.

\section{\normalsize \bf  Conclusions}
We have constructed a phenomenological model in which the cosmological
constant is considered as a step function depending on the epoch. Also
ordinary standard matter is taken into consideration as radiation at the
beginning and as dust after the equivalence. The model is exactly solvable
and allows to implement an inflationary epoch after which we found
substantial agreement with the most recent observational data concerning the
values of $\Omega$, the age of the universe, the CMBR temperature at
equivalence and today. The construction can be perfectly compatible with
such models which call for an amount of barionic matter, cold dark matter
and cosmological constant in order to explain cosmological dynamics and
large scale structure formation after an inflationary expansion
\bib{starobinsky}. We have to remark that this is a toy model in which a
sharp transition between the two values of $\Lambda$ is invoked but it
justifies how a time-dependent cosmological constant could affect early and
present cosmological dynamics and, in some sense, be in agreement with
observational data. That is this analysis confirms how important it could be 
to specify the concept of ``cosmological constant'' in a wider way and in a
more general context, such as, for example, that of the nonminimally coupled
scalar-tensor theories, where also scalar field(s) can be considered in
dynamics and, in general, be nonminimally coupled with geometry
\bib{noether}.

It is noteworthy that the technique we used to solve (\ref{1})-(\ref{3}) of
multipling the first order Einstein equation for an opportune factor and
adding what obtained to the second order Einstein equation to get an
equation in which it does not appear explicitely the term in $\rho_m$ can be
used also in  more general cases where a scalar field energy density is
present together with that of matter. The case we have considered is a
particular case of constant potential and zero initial conditions for the
scalar field dynamics.

Furthermore we have to say that the problem of  ``graceful exit'' has to be
taken into consideration; that is we have to consider what kind of realistic
inflationary model could implement the above dynamics and allow the
requested $\Lambda$ transition.

As concluding remark we consider with more attention the non completely
satisfactory compatibility expressed by (\ref{46}); actually, giving to $h$
the value $h=0.7$ compatible with the data, the square of the relative
difference reaches its minimum for a negative value of $t_{eq}$. Thus the
transition from radiation to matter would occur before the initial
singularity, which means that to have full compatibility relatively to the
matter energy density, a radiation dominated era does not find space. This
suggests an explanation of the disagreement concerning the matter in
presence of radiation, because we have totally neglected its presence after
$t_{eq}$ but we have taken it into account in the comparison with the data.
The order of magnitude of the disagreement is in fact less than the order of
magnitude of the ratio between the present energy density of the radiation
and that of the matter (see eqs. (\ref{46}) and (\ref{39}), (\ref{42})). It
is reasonable to think, in order to solve these discrepancies, that a model
with non zero initial conditions on the scalar field (see \bib{marek}) can
be revisited with the phenomenological approach used in this paper. This
will be one of the further developments of our future research.

\vspace{5. mm}

\begin{centerline}
{\bf Acknowledgements}
\end{centerline}
One of us (A. A. M.) likes to thank S.Matarrese for the very useful
discussions had on this topic.
\vspace{5. mm}

\begin{centerline}
{\bf References}
\end{centerline}

\begin{enumerate}
\item \label{fuku-taka}
M. Fukugita, F. Takahara, K. Yamashita, and Y. Yoshii, \apj {\bf 361} (1990) 
L1.
\item\label{ostri}
J. P. Ostriker and P. J. Steinhardt, \nat {\bf 377} (1995) 600.
\item\label{carrol}
S.M. Carrol, W. H. Press, and E. L. Turner, \araa {\bf 30} (1992) 499.
\item\label{wein}
S. Weinberg, \rmp {\bf 61} (1989) 1.
\item \label{fuku-hoga}
M. Fukugita, C. J. Hogan, and P. J. Peebles, \nat {\bf 366} (1993) 309.
\item\label{freed}
W. L. Freedman {\it et al.} \nat {\bf 371} (1994) 757.
\item \label{fuku-futa}
M. Fukugita, T. Futamase, M. Kasai, and E. L. Turner \apj {\bf 393} (1992) 
3.
\item\label{guth}
A. H. Guth, \pr {\bf B 23} (1981) 347.
\item\label{linde}
A. D. Linde \pl {\bf 108 B} (1982) 389.
\item\label{peebles}
P. J. E. Peebles, {\it Priciples of Physical Cosmology}, Princeton Univ. 
Press, Princeton (1993).
\item\label{ellis}
M. S. Madsen and G. F. R. Ellis, \mnras {\bf 234} (1988) 67.
\item\label{wright}
E. L. Wright, to appear in {\it International School of Physics ``Enrico 
Fermi'' Course CXXXII: Dark Matter in the Universe}, eds. S. Bonometto, 
J. R. Primack, and A. Provenzale, Varenna (1995).
\item\label{pier}
M. J. Pierce {\it et al.}, \nat {\bf 371} (1994) 385.
\item\label{efsta}
G. Efstathiou, W. J. Sutherland, and S. J. Maddox, \nat {\bf 348} (1990) 705.
\item\label{dekel}
A. Dekel, \araa {\bf 32} (1994) 371.
\item\label{prim}
J. R. Primack, to appear in {\it International School of Physics ``Enrico 
Fermi'' Course CXXXII: Dark Matter in the Universe}, eds. S. Bonometto, 
J. R. Primack, and A. Provenzale, Varenna (1995).
\item\label{kolb}
E. W. Kolb and M. S. Turner, \arns {\bf 33} (1983) 645
\item\label{starobinsky}
A.A. Starobinsky, in {\it Cosmoparticle Physics 1}, eds. M.Yu. Khlopov 
{\it et al.} Edition Frontier (1996).
\item\label{noether}
S. Capozziello, R.de Ritis, C. Rubano, and P. Scudellaro, to appear in {\it 
La Riv. del Nuovo Cim.} (1996).
\item\label{marek}
M. Demianski, R. de Ritis, C. Rubano, and P. Scudellaro, \pr {\bf D 46} 
(1992), 1391.

\end{enumerate}

\vfill

\end{document}